\documentclass[pra,twocolumn,superscriptaddress,longbibliography]{revtex4-2}
\usepackage[colorlinks=true, citecolor=blue, urlcolor=blue, linkcolor=blue ]{hyperref}
\usepackage{empheq,amssymb,orcidlink,braket}

\begin{document}
\title{Quantum router of silicon-vacancy centers via a diamond waveguide}
\author{Wen-Jie Zhang}
\affiliation{Key Laboratory of Quantum Theory and Applications of MoE, Lanzhou Center for Theoretical Physics, Gansu Provincial Research Center for Basic Disciplines of Quantum Physics, Key Laboratory of Theoretical Physics of Gansu Province, Lanzhou University, Lanzhou 730000, China}
\author{Xi Yan}
\affiliation{Key Laboratory of Quantum Theory and Applications of MoE, Lanzhou Center for Theoretical Physics, Gansu Provincial Research Center for Basic Disciplines of Quantum Physics, Key Laboratory of Theoretical Physics of Gansu Province, Lanzhou University, Lanzhou 730000, China}
\author{Jun-Hong An\orcidlink{0000-0002-3475-0729}}\email{anjhong@lzu.edu.cn}
\affiliation{Key Laboratory of Quantum Theory and Applications of MoE, Lanzhou Center for Theoretical Physics, Gansu Provincial Research Center for Basic Disciplines of Quantum Physics, Key Laboratory of Theoretical Physics of Gansu Province, Lanzhou University, Lanzhou 730000, China}
	
\begin{abstract}
As a key component of quantum networks, the quantum router distributes quantum information among different quantum nodes. The silicon-vacancy (SiV) center in diamond offers a promising platform for quantum technology due to its strong strain-induced coupling with phonons. However, the development of a practical quantum router faces the challenges of achieving long-range entanglement and suppressing decoherence. Here, we propose a non-Markovian quantum router based on a diamond waveguide embedded with an array of SiV centers as the quantum nodes. Unlike conventional channel-switching methods, our design enables parallel quantum-state transfer from a single input node to multiple target nodes, analogous to a classical WiFi router. We demonstrate that persistent entanglement and suppressed decoherence of the SiV centers over long distances are achievable when bound states are present in the energy spectrum of the total system formed by the SiV centers and the phonon waveguide. Our scheme enriches the implementation of quantum routing and prompts the development of solid-state quantum networks.
\end{abstract}
\maketitle
	
\section{Introduction}
Quantum networks formed by interconnecting multiple quantum channels between quantum nodes serve as a significant constituent for the realization of quantum communication, quantum computing, and quantum sensing \cite{PhysRevApplied.6.040001,Kimble2008,PhysRevA.98.030302,Gisin2007,Chen2021,Guo2019,Main2025}. A central challenge in the development of practical quantum networks lies in achieving efficient internode quantum information transfer, a task in which quantum routers play an essential role \cite{8068178,10269671,PhysRevA.111.042423,Lee2022,PhysRevLett.105.260501}. Extensive research on quantum routing has been conducted on various physical platforms, such as superconducting circuits \cite{PhysRevApplied.15.014049,PhysRevA.97.052315,PhysRevApplied.6.024009}, cavity quantum electrodynamics (QED) systems \cite{PhysRevResearch.2.043048,PhysRevA.97.023801,PhysRevX.3.031013}, and atomic and optical systems \cite{PhysRevLett.106.053901,PhysRevA.87.062333,doi:10.1126/science.1254699}. While the above platforms have advanced quantum routing, they still face practical limitations such as on‑chip dissipation, miniaturization hurdles, and dependence on complex external controls. Recently, the outstanding performance of diamond-based color centers in building phonon networks has motivated us to explore the implementation of quantum routing on this promising platform \cite{PhysRevLett.120.213603,PhysRevX.13.041037}.

Diamond color centers, such as the germanium-vacancy (GeV), nitrogen-vacancy (NV), tin-vacancy (SnV), and silicon-vacancy (SiV) systems, play significant roles in quantum science and technology \cite{PhysRevLett.118.223603,PhysRevA.98.052346,Doherty2013,doi:10.1126/science.1131871,PhysRevLett.113.263601,PhysRevLett.122.063601,PhysRevX.13.041037,PhysRevApplied.15.064010,becker_ultrafast_2016}. Although NV centers are widely employed and exhibit outstanding coherence properties \cite{doi:10.1126/science.1231364,PhysRevX.6.041060,PhysRevA.97.052303,PhysRevLett.108.043604,balasubramanian_ultralong_2009,PhysRevA.96.063810,PhysRevLett.108.143601,PhysRevLett.117.015502,Maze_2011}, their intrinsically weak coupling to diamond phonon modes limits their performance in strong-coupling regimes. In contrast, centers such as SiV and SnV feature strain-mediated phonon couplings that exceed those of NV centers by several orders of magnitude \cite{PhysRevB.97.205444,Jahnke_2015,sohn_controlling_2018,PhysRevX.13.041037}, establishing them as leading candidates among solid-state quantum platforms \cite{PhysRevApplied.11.024073,Lekavicius:19,PhysRevLett.107.235502,PhysRevX.13.041037}. This characteristic offers us a different approach to phonon-mediated quantum manipulation. Based on this advantage, numerous studies have been conducted, including the development of phonon-based quantum networks \cite{PhysRevLett.120.213603}, the generation of spin-squeezed states \cite{PhysRevA.103.013709,WOS:000541886200002,Ren:24,PhysRevA.110.052610,PhysRevA.101.042313}, and quantum simulation of the Su-Schrieffer-Heeger models \cite{PhysRevResearch.2.013121,PhysRevResearch.3.013025,PhysRevResearch.4.023077}. The spin-phonon coupling properties and scalability of diamond waveguides endow the coupled system of SiV centers and phonons as a natural platform for solid-state quantum routers. However, three fundamental challenges remain in the realization of the quantum routing function \cite{8068178}. First, the generated entanglement decreases dramatically with increasing the distance among the quantum nodes. Second, the inevitable environmental noise in diamond waveguides causes decoherence on both quantum entanglement and state transfer of the SiV centers. Third, conventional quantum routers typically operate as end-to-end switches \cite{PhysRevLett.111.103604,PhysRevA.92.063836,PhysRevA.105.013711,PhysRevA.98.063809,PhysRevA.99.033827,PhysRevA.109.053707,PhysRevA.89.013805}, which makes it hard to support a parallel state transfer to multiple nodes. Thus, achieving robust long-range entanglement, suppressing decoherence during state transmission, and overcoming the end‑to‑end limitation of conventional quantum routers are highly desired.

In this work, to overcome these limitations, we propose a quantum router operating on a principle similar to that of a classical WiFi. Enabling a simultaneous transfer of quantum states from a single source node to multiple target nodes, our scheme works in the non-Markovian dynamics of an array of SiV centers interacting with the phonon modes in a one-dimensional diamond waveguide. We find that the strain-induced strong spin-phonon coupling in the diamond waveguide drives the hybrid SiV-phonon system to form multiple bound states, whose formation makes the long-distance entanglement among multiple nodes persistent and the multiplexed quantum-state transfer via the waveguide robust to the decoherence. Our result efficiently solves these key challenges in quantum routing. It not only enriches the implementation approaches of quantum routing, but also provides an effective pathway for scalable noise-resistant quantum networks.

This paper is organized as follows. In Sec. \ref{1}, we show the system consisting of an array of SiV centers coupled to a one-dimensional diamond waveguide. In Sec. \ref{2}, we derive the energy spectrum of the total system and investigate its non-Markovian dynamics of the SiV centers. A comprehensive discussion is presented on the entanglement and state transfer dynamics in quantum router containing two and three SiV centers. Finally, we provide the discussions on the experimental realization of our proposal and summarize the main conclusions in Sec. \ref{3}.

\begin{figure}
	\includegraphics[width=1\columnwidth]{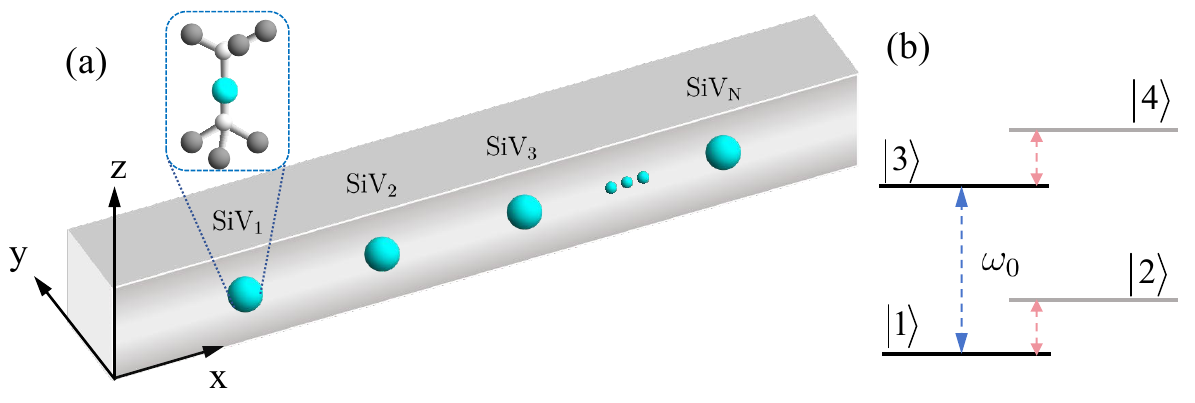}
	\caption{  (a) Schematic of $N$ SiV centers embedded in a one-dimensional phonon waveguide. (b) In a magnetic field, the SiV center is seen as a four-level system, whose sublevels are $\ket{i}$ ($i=1,\ldots,4$). Transition between sublevels with different orbital degrees of freedom of $\ket{1}\leftrightarrow \ket{3}$ or $\ket{2} \leftrightarrow \ket{4}$  has transition frequency $\omega_{0}$. At zero magnetic field, there are two orbital branches in the ground state of SiV center, which can be viewed as a two-level system denoted by $ \ket{1} $ and $ \ket{3} $. }\label{FIG1}
\end{figure} 
	
\section{System}\label{1}
We consider a system consisting of an array of $N$ SiV centers coupled to a one-dimensional diamond waveguide, as illustrated in Fig. \ref{FIG1}(a). Each SiV center is formed by a silicon atom and a split vacancy that replaces two adjacent carbon atoms in the diamond lattice \cite{PhysRevLett.112.036405}. The electronic ground state of the SiV center comprises an unpaired hole with spin $S=1/2$, occupying one of two degenerate orbital states $|e_x\rangle$ and $|e_y\rangle$. Under the combined effects of spin-orbit coupling and a weak Jahn-Teller interaction, the four resulting states are split into two doublets: $\{|1\rangle\simeq|e_{-},\downarrow\rangle, |2\rangle\simeq|e_{+},\uparrow\rangle\}$ and $\{|3\rangle\simeq|e_{+},\downarrow\rangle, |4\rangle\simeq|e_{-},\uparrow\rangle\}$, where $\left|e_{\pm}\right\rangle = \left(\left|e_x\right\rangle \pm i\left|e_y\right\rangle\right)/\sqrt{2}$ are eigenstates of the orbital angular momentum operator $\hat{L}_z$, satisfying $\hat{L}_z \left| e_{\pm} \right\rangle = \pm \hbar \left| e_{\pm} \right\rangle$. The energy gap between these doublets is $\Delta/2\pi \simeq 46\,\mathrm{GHz}$ \cite{PhysRevLett.112.036405}. The one-dimensional diamond phononic waveguide has the dimensions $L$, $a$, and $b$, where $L\gg a, b$. In the case of a linear isotropic medium, the phonon modes in the waveguide are described as elastic waves characterized by a displacement field ${\bf u}({\bf r})$. Under the periodic boundary condition, the displacement field is quantized as \cite{PhysRevLett.120.213603}
\begin{equation}\label{e4}
\hat{\bf u}({\bf r}) = \sum_{n, k} \Big(\frac{\hbar}{2 \varrho V \omega_{n, k}}\Big)^{1\over 2} {\bf u}_{n, k}^{\perp}(y, z) ( \hat{a}_{n, k} e^{i k x} + \text{H.c.} ),
\end{equation} 
where $V=Lab$ is the volume of the waveguide, $\varrho$ is the material density, and ${\bf u}_{n, k}^{\perp}(y, z)$ is the transverse profile of the displacement field, with $k$ being the wave vector and $n$ being the branch index. Then the Hamiltonian of the phonon is $\hat{H}_{\text{ph}}=\hbar\sum_{n,k}\omega_{n,k}\hat{a}_{n,k}^{\dagger} \hat{a}_{n,k}$, where $\hat{a}_{n,k}$ is the annihilation operator of the $k$th phonon mode with frequency $\omega_{n,k}$ \cite{stroscio_acoustic_1996,PhysRevB.94.214115,PhysRevB.97.205444}. The frequencies of the phonons in the waveguides for a small cross section and a longe distance $L$ become continuous. 

The electronic states of the SiV centers embedded in the waveguide are modulated by the compressive modes arising from lattice distortions in the phononic waveguide, leading to a strain-mediated coupling between the phonons and the orbital degrees of freedom of the SiV centers \cite{PhysRevB.94.214115}. Under the rotating-wave approximation, the resulting strain-coupling Hamiltonian between the phonons and the array of $N$ SiV centers is $\hat{H}_{\text{strain}}=\hbar\sum_{j,n,k}(g_{k,j}^{(n)}\hat{J}_+^j\hat{a}_{n,k}+\text{H.c.})$, where $\hat{J}_-^j= \ket{1}_j\bra{3}+\ket{2}_j\bra{4}$ is the spin-conserving lowering operator. The coupling strength is $g_{k,j}^{(n)} = d (\frac{\hbar k^{2}}{2 \varrho lS \omega_{n, k}})^{1/2} \xi_{n, k}\left(y_j, z_j\right)e^{ikx_j}$, where $x_j$ is the position of the $j\text{th}$ SiV center in the waveguide, $d/2\pi \sim\mathrm{PHz}$ is the strain sensitivity, and $S$ is the cross-sectional area of the waveguide  \cite{sohn_controlling_2018,PhysRevB.110.045419}. The dimensionless function $\xi_{n,k}(y_j, z_j)$ describes the spatial strain profile and depends on the local strain tensor $\epsilon^{xy}_{n,k}(x_j,y_j)={1\over 2}[\partial_{y}u^{\perp z}_{n,k}(y,z)+\partial_z u_{n,k}^{\perp,y}(y,z)]|_{y=y_j,z=z_j}$. For a homogeneous compression mode, a widely used approximation is $\xi_{n,k}(y_j, z_j) \simeq 1$ \cite{PhysRevLett.120.213603}. Furthermore, when the transverse cross-sectional area of the phonon waveguide is sufficiently small compared to the characteristic wavelength of phonons, the coupling between transverse phonon modes and the SiV centers can be neglected. It suffices to consider only the longitudinal phonon modes in the SiV-phonon coupling. Therefore, the branch index $n$ can be omitted. In the absence of a magnetic field, each SiV center can be treated as an effective two-level system, where the degenerate doublet $\{|1\rangle,|2\rangle \}$ constitutes the ground state and $\{|3\rangle,|4\rangle \}$ constitutes the excited state \cite{https://doi.org/10.1002/qute.202100074}. Then the Hamiltonian of the total system becomes 
\begin{equation}
\hat{H} = \hbar\omega_0 \sum_{j = 1}^{N} \hat{\sigma}_j^\dagger\hat{\sigma}_j +\hbar \sum_{k} \big[ \omega_k \hat{a}_k^\dagger \hat{a}_k  +\sum_{j = 1}^{N} ( g_{k,j} \hat{a}_k^\dagger \hat{\sigma}_j + \text{H.c.} ) \big], \label{HH}
\end{equation}
where $\hat{\sigma}_j=\ket{1}_j\bra{3}$ is the transition operator from $|3\rangle$ to $|1\rangle$ with the transition frequency $\omega_{0}$. The transition frequency is widely tunable from $46 \,\mathrm{GHz}$ to $1.2 \,\mathrm{THz}$ via static strain control \cite{sohn_controlling_2018}. The dispersion relation of the phonons is taken to be linear as $\omega_{k}=vk$, where $v$ is the group velocity of the longitudinal phonon modes \cite{PhysRevA.101.042313,PhysRevLett.120.213603}. 

\section{Non-Markovian Quantum Routing }\label{2}
We propose to use the array of $N$ SiV centers coupled to the one-dimensional diamond waveguide to realize a quantum router. The first SiV center acts as the input port for the router. Via the waveguide, we hope to realize the simultaneous transfer of quantum information to other SiV centers acting as the output ports of the router. 

We prepare only the first SiV center in the excited state and all the phonon modes in the vacuum state initially, i.e., $|\Psi(0)\rangle=\hat{\sigma}_1^\dag|G,\{0_k\}\rangle$. Here, $|G\rangle$ denotes that all the SiV centers are in their ground state and $|\{0_k\}\rangle$ is the vacuum state of the phonon modes in the waveguide. This vacuum-state assumption is physically well justified in the low-temperature regime $T \ll \hbar\Delta / k_{B} \approx 2.2\mathrm{~K}$, where the thermal occupation number is entirely negligible \cite{PhysRevLett.120.213603}. It is straightforward to verify that the total excitation number $\hat{\mathcal{N}} = \sum_{j} \hat{\sigma}_{j}^{\dagger} \hat{\sigma}_{j} + \sum_{k} \hat{a}_{k}^{\dagger} \hat{a}_{k}$ is conserved due to $[\hat{H}, \hat{\mathcal{N}}] = 0$. Therefore, the time evolved state is expressed as
\begin{equation}
|\Psi(t)\rangle =[\sum_j c_j(t) \hat{\sigma}_j^\dagger + \sum_k r_k(t) \hat{a}_k^\dagger ] |G; \{0_k\}\rangle,
\end{equation}
where $c_j(t)$ is the excited-state probability amplitudes of the $j$th SiV center, $r_k(t)$ is the probability amplitude of the phonon field with only one phonon in the $k\mathrm{th}$ mode. From the Schr\"{o}dinger equation, we derive 
\begin{eqnarray}
i\dot{c}_j(t) &=& \omega_0 c_j(t) + \sum_k g_{jk} r_k(t),\label{S3}\\ 
i\dot{r}_k(t) &=& \omega_k r_k(t) + \sum_{l=1}^N g_{lk}^* c_l(t)\label{S4}.
\end{eqnarray}
The substitution of the solution of Eq. \eqref{S4} as $r_k(t) = -i \sum_l \int_0^t d\tau \, g_{lk}^* e^{-i\omega_k (t-\tau)} c_l(\tau)$ under the initial condition $r_k(0)=0$ into Eq. \eqref{S3} leads to  
\begin{equation}
\dot{c}_j(t) + i\omega_0 c_j(t) + \sum_{l=1}^{N} \int_{0}^{t} h_{jl}(t - \tau) c_l(\tau) d\tau = 0,\label{ct1}
\end{equation}
where $h_{jl}(t-\tau)=\int_0^\infty d\omega J_{jl}(\omega)e^{-i\omega(t-\tau)}$ is the correlation function of the phonon mode and $J_{jl}(\omega)=2\pi\sum_kg_{jk}g_{lk}^*\delta(\omega-\omega_k)$ denotes the correlated spectral density between the $j\text{th}$ and $l\text{th}$ SiV centers. The existence of such a correlation shows that the phonons in the diamond waveguide effectively mediate indirect couplings among the spatially separated SiV centers. This phonon-mediated coupling naturally implements an intrinsic quantum routing mechanism by establishing multiplexed coherent channels among distant SiV centers. The spectral density function is explicitly derived as
\begin{equation}
J_{jl}(\omega)={d^2 \hbar \omega e^{-\frac{\omega}{\omega_c}}}\cos [{\omega (x_{j} - x_l)}/{v}] /(\varrho S v^3 ),\label{J}
\end{equation}
where $\omega_c$ is a cutoff frequency to remove the integration divergence in the correlation function. It is observed that $J_{11}(\omega)=\dots=J_{NN}(\omega)$ and $J_{jl}(\omega)=J_{lj}(\omega)$. Therefore, $J_{jl}$ satisfies the property $J_{jl}(\omega)=J_{|j-l|}(\omega)$. Accordingly, Eq. \eqref{ct1} is rewritten in column vector form as
\begin{equation}
\dot{\textbf{c}}(t)+i\omega_0\textbf{c}(t)+\int_0^td\tau\mathbf{h}(t-\tau)\mathbf{c}(\tau)=0,\label{c}
\end{equation}
where $\mathbf{c}(t)=\left(
  \begin{array}{ccccc}
    c_1(t) & \cdots & c_N(t) \\
  \end{array}
\right)^\text{T}$, $\mathbf{h}(t-\tau)=\int_0^\infty d\omega e^{-i\omega(t-\tau)}\mathbf{J}(\omega)$ is an $N\times N$ matrix with elements $h_{jl}(t-\tau)$, and $\mathbf{J}(\omega)$ is a spectral density matrix with elements $J_{jl}(\omega)$.  

The convolution in Eq. \eqref{c} reflects the non-Markovian memory effect. The effect is especially pronounced in SiV-diamond systems due to the large strain sensitivity of SiV centers and the resulting enhancement of phonon-matter couplings. However, a widely used treatment in the literature is to neglect this effect \cite{PhysRevLett.120.213603,PhysRevB.94.214115,PhysRevA.103.013709,PhysRevA.101.042313,WOS:000541886200002}. Under the Markov approximation, the memory effect is neglected by replacing $\mathbf{c}(\tau)$ with $\mathbf{c}(t)$ and extending the upper limit of the time integral to infinity. The resulting approximate solution is $\mathbf{c}^{\text{MA}}(t) = \exp\left[-(\boldsymbol{\Gamma} + i\omega_0 + i\boldsymbol{\Delta})t\right]\mathbf{c}(0)$, where $\boldsymbol{\Gamma} = \pi \mathbf{J}(\omega_0)$ and $\boldsymbol{\Delta}= \mathcal{P} \int \frac{ \mathbf{J}(\omega_0) d\omega}{\omega_0 - \omega}$, with $\mathcal{P}$ denoting the Cauchy principal value. $|\mathbf{c}^{\text{MA}}(t)|^2$ decays exponentially to zero due to the positive definiteness of ${\pmb \Gamma}$. It implies that the destructive influence of the Markovian decoherence inhibits the transfer of quantum information among the SiV centers and the implementation of quantum router by the diamond waveguide. 

Going beyond the Markov approximation, we exactly study the correlation dynamics of the $N$ SiV centers mediated by the diamond waveguide. Although the exact solution of Eq. \eqref{c} is only obtained by numerical calculation, its long-time behavior can be analytically derived via the Laplace transformation. Specifically, Eq. \eqref{c} is transformed into
$[s + i\omega_0 + \tilde{\mathbf{h}}(s)]\tilde{\mathbf{c}}(s) = \mathbf{c}(0)$, where $\tilde{\bf h}(s)=\int_0^\infty d\omega{{\bf J}(\omega)\over s+i\omega}$ is the Laplace transform of ${\bf h}(t-\tau)$. Using the Jordan decomposition $\mathbf{K}(s) = \mathbf{A}_s^{-1} \tilde{\mathbf{h}}(s) \mathbf{A}_s = \text{diag}[K_1(s), \cdots, K_N(s)],$ we obtain $\tilde{\mathbf{c}}(s) = \mathbf{A}_s \left[ s + i\omega_0 + \mathbf{K}(s) \right]^{-1} \mathbf{A}_s^{-1} \mathbf{c}(0)$. The time-domain solution \(\mathbf{c}(t) \) is obtained after performing the inverse Laplace transform to \(\tilde{\mathbf{c}}(s) \), which requires finding the poles of \(\tilde{\mathbf{c}}(s) \) via the equation
\begin{equation}
Y_j(\varpi) \equiv \omega_0 - iK_j(-i\varpi) = \varpi,~(\varpi=i s). \label{Y}
\end{equation}It is interesting to find that the root $\varpi$ multiplied by $\hbar$ corresponds exactly to the eigenenergies of Eq. \eqref{HH}. To prove this, we expand its eigenstate as $| \Phi \rangle =[\sum_{j}m_j\hat{\sigma}_j^\dagger + \sum_{k} n_k \hat{a}_k^\dagger ] |G; \{0_k\}\rangle $. From $\hat{H}\ket{\Phi} = E\ket{\Phi}$, we obtain
\begin{eqnarray}
(E/\hbar-\omega_0)m_j&=&\sum_kg_{jk}n_k,\label{s1}\\
(E/\hbar-\omega_k)n_k&=&\sum_{l=1}^Ng_{lk}^*m_l.\label{s2}
\end{eqnarray}
Substituting the solution $n_k=\sum_{l=1}^Ng_{lk}^*m_l/(E/\hbar-\omega_k)$ from Eq. \eqref{s2} into Eq. \eqref{s1} yields $(E/\hbar-\omega_0)m_j=-i\sum_{l=1}^N\tilde{h}_{jl}(-iE/\hbar)m_l$, which is concisely expressed in a matrix form as
\begin{equation}
[E/\hbar-\omega_0+i\tilde{\textbf{h}}(-iE/\hbar)]\textbf{m}=0,\label{smeige}
\end{equation}where $\textbf{m}=\left(
  \begin{array}{ccccc}
    m_1 & \cdots & m_N \\
  \end{array}
\right)^\text{T}$. Through the Jordan decomposition of $\tilde{\bf h}$, we convert Eq. \eqref{smeige} into
$E/\hbar-\omega_0+iK_j(-iE/\hbar)=0$, which determines the eigenenergies $E$ of the total system and matches Eq. \eqref{Y} exactly. Therefore, Eq. \eqref{Y} not only governs the dynamical evolution of $\mathbf{c}(t)$ but also determines the eigenenergies of the total system. This implies that the dynamics described by $\mathbf{c}(t)$ is intrinsically determined by the energy-spectrum properties of the total system. 

Equation \eqref{Y} has two types of solutions. For \( \varpi> 0\), the function \(Y_j(\varpi)\) is ill-defined and exhibits rapid jump between $\pm\infty$. Thus, Eq. \eqref{Y} has an infinite number of roots in the regime $\varpi>0$, which form a continuous energy band. In contrast, for \(\varpi < 0\), each \(Y_j(\varpi)\) is a monotonic decreasing function. Thus, we always have an isolated root \(\varpi_j^b\) in the regime $\varpi<0$ provided \(Y_j(0)<0\). The eigenstates associated with these discrete eigenenergies \( \hbar \varpi_j^{\text{b}} \) are dubbed bound states. The formation of such bound states plays a crucial role in the non-Markovian dynamics \cite{PhysRevLett.132.090401,yang_long-range_2025}. By applying the Cauchy's residual theorem and contour integration techniques \cite{PhysRevA.103.L010601,PhysRevB.106.115427}, we obtain
\begin{equation}
\mathbf{c}(t) = \mathbf{Z}(t) + \int_{0}^{\infty} \frac{d\varpi}{2\pi} [\tilde{\mathbf{c}}(0^+ - i\varpi) - \tilde{\mathbf{c}}(0^- - i\varpi)] e^{-i\varpi t},\label{ct}
\end{equation}
where ${\bf Z}(t) =\sum_{j=1}^{N} \text{Res}[\tilde{\mathbf{c}}(-i\varpi_j^b)] e^{-i\varpi_j^b t}$ and $\text{Res}[\tilde{\mathbf{c}}(-i\varpi_j^b)]$ denotes the residue contributed by the $j\text{th}$ bound state. The second term originates from the continuous-energy band. This term oscillates with a continuously varying frequency $\varpi$ and decays to zero in the long-time limit due to out-of-phase interference. Therefore, if no bound states exist, $\lim_{t \to \infty} \mathbf{c}(t) = \mathbf{0}$, indicating the complete decoherence and failure of quantum-information transfer. In contrast, the formation of bound states leads to $\lim_{t \to \infty} \mathbf{c}(t) = \mathbf{Z}(t)$, signifying the suppression of decoherence and a persistent quantum-information transfer among different SiV centers. Being absent in the Markov approximation, this behavior is a distinctive feature of non-Markovian dynamics. Being enabled by this mechanism, we realize a quantum router based on diamond waveguides. Analogous to conventional WiFi networks but operating through non-Markovian dynamics, this quantum router utilizes bound-state formation to establish stable entanglement among the SiV center nodes. As we will see later, it enables a parallel one-to-many quantum state transfer within the network.

First, we consider the case of $N=2$ under the initial state $|\Psi(0)\rangle = \frac{1}{\sqrt{2}}(|0\rangle_1 +|1\rangle_1 )|0\rangle_2 $. The spectral density matrix is $\mathbf{J}(\omega) = \begin{pmatrix} J_0(\omega) & J_1(\omega) \\ J_1(\omega) & J_0(\omega) \end{pmatrix}$. It is easy to derive $\mathbf{K}(s)={\rm diag}[\tilde{h}_{0}(s)+\tilde{h}_{1}(s), \tilde{h}_{0}(s)-\tilde{h}_{1}(s)]$ and $\mathbf{A}_s = \begin{pmatrix}
1 & 1 \\
1 & -1
\end{pmatrix}$. Thus, we have $\tilde{\mathbf{c}}(s) = \mathbf{Q} \left[ s + i \omega_{0} + \mathbf{K}(s) \right]^{-1} \begin{pmatrix} 1 & 1 \end{pmatrix}^{\mathrm{T}}$, where $\mathbf{Q} = \begin{pmatrix} 1/2 & 1/2 \\ 1/2 & -1/2 \end{pmatrix}$. Next, we need to make the inverse Laplace transform to obtain $\mathbf{c}(t)$. The contribution to the inverse Laplace transform of $\left[ s + i \omega_{0} + \mathbf{K}(s) \right]^{-1}$ by the $l\mathrm{th}$ bound state with frequency $\omega_l^b$ is given by the residue $\lim\limits_{s \to -i\varpi_{l}^{b}}(s + i\varpi_{l}^{b}) e^{st} [s + i\omega_{0} + K_{j}(s)]^{-1}= Z_{l} e^{-i\varpi_{l}^{b} t} \delta_{l, j}$, where $Z_l= \left.[1 + \partial_s K_l(s)]^{-1}\right|_{s \to -i\varpi_l^b}$. Since the steady-state form of $\mathbf{c}(t)$ obtained by the inverse Laplace transform of $\tilde{\mathbf{c}}(s)$ is governed by the residues of the bound states, we have
\begin{equation}
c_j(\infty) = \sum_{l=1}^{2} Q_{jl} Z_l e^{-i\varpi_l^bt}.\label{c2}
\end{equation}
\begin{figure}
\includegraphics[width=1.0\columnwidth]{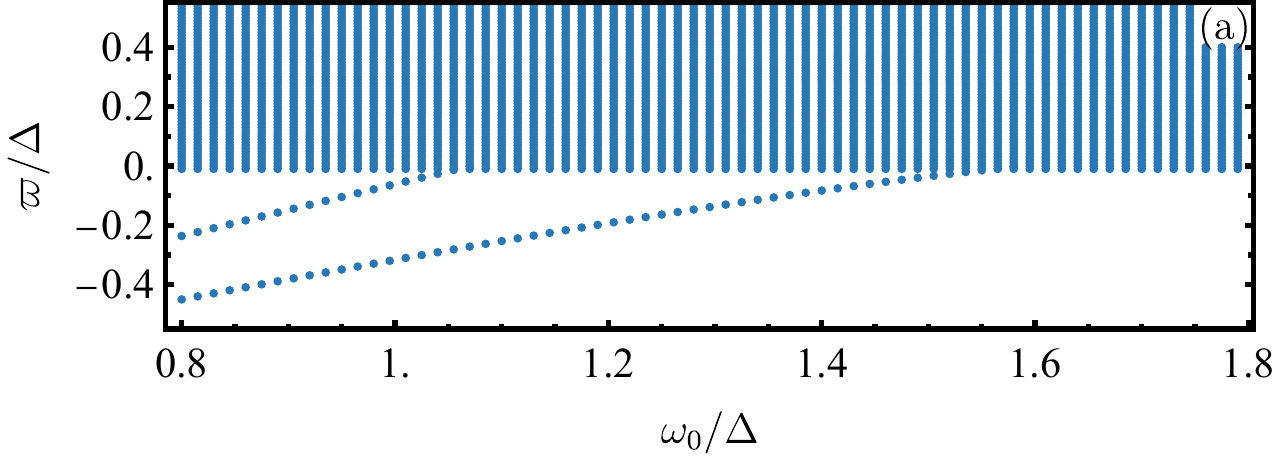}
\hspace*{3pt}\includegraphics[width=0.96\columnwidth]{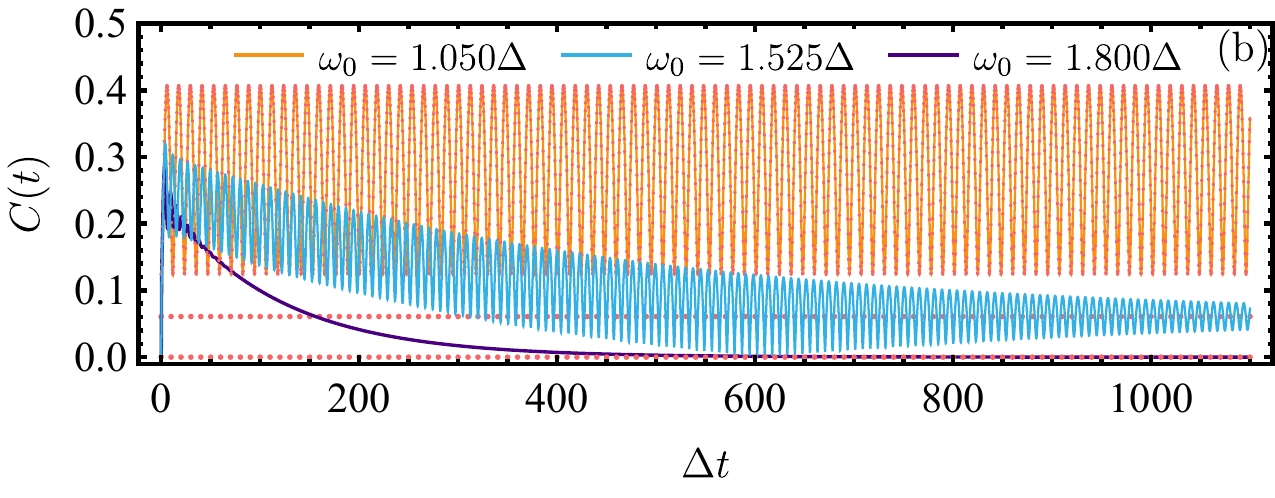}
\hspace*{5pt}\includegraphics[width=0.985\columnwidth]{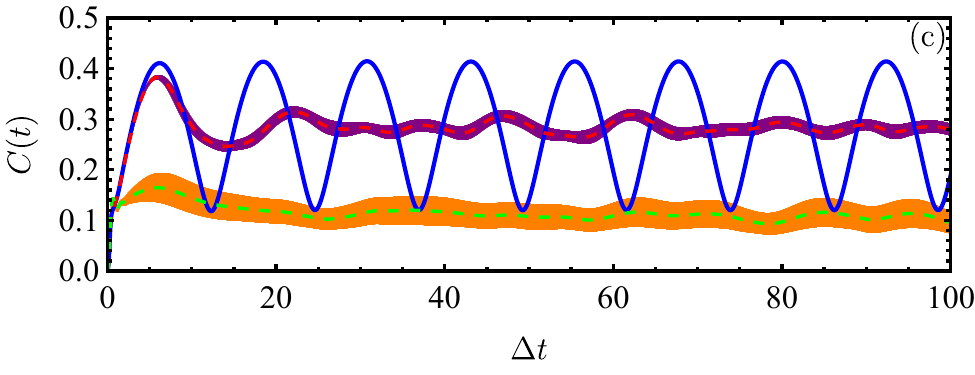}
\caption{(a) Energy spectrum of the total system obtained by numerically solving Eq. \eqref{Y}. (b) Time evolution of the concurrence \(C(t)\) obtained by numerically solving Eq. \eqref{c} for different values of \(\omega_0\). The orange, blue, and purple curves correspond to the cases with two, one, and zero bound states, respectively. The red dotted lines represent the analytical results of Eq. \eqref{c2}, showing a good agreement with the numerical results. (c) Evolution of concurrence $C(t)$ when a random fluctuation $\chi$ is added on the frequency $\omega_0=\Delta$ of the SiV centers. The red and green dashed lines are the average values obtained by averaging over 100 random configurations when $\chi=[-0.5\Delta,0.5\Delta]$ and $[-\Delta,5\Delta]$. The purple and orange regions are their respective variations. The blue solid lines are the fluctuation-free results. We use $\omega_c=7\Delta$, $S=100\,\mathrm{nm} ^2$, $d/2\pi \approx 4\,\mathrm{PHz}$, $\delta x\equiv |x_1-x_2|=10\,\mathrm{nm}$, \( v \approx 1 \times 10^4 \, \text{m/s}\), and $N=2$.}\label{F2}
\end{figure}
The bipartite entanglement between the SiV centers in a state \(\rho\) is quantified by the concurrence $C = \max(0, \lambda_1 - \lambda_2 - \lambda_3 - \lambda_4)$, where \(\lambda_i\) are the square roots of the eigenvalues of the matrix $\rho(\hat{\sigma}_y \otimes \hat{\sigma}_y) \rho^*(\hat{\sigma}_y \otimes \hat{\sigma}_y)$ in the decreasing order   \cite{PhysRevLett.80.2245}. In our case, we have 
\begin{equation}
C(t)  = \sqrt{2} |c_1(t) c_2^*(t)|,
\end{equation}
which, according to Eq. \eqref{c2}, is obviously not zero in the long-time limit as long as the bound states are formed. We are interested in examining how the quantum information encoded in the first SiV center transfers to the second one as $\frac{1}{\sqrt{2}} (|0\rangle_1 +|1\rangle_1 )|0\rangle_2\rightarrow |0\rangle_1  \frac{1}{\sqrt{2}} (|0\rangle_2 + |1\rangle_2)$. The fidelity of this state transfer is given by 
\begin{equation}
F(t)=| \langle \Phi_{\text{target}} | \Psi(t) \rangle |^2=|1+  c_2(t) |^2/4,\label{fdtl}
\end{equation}
where the target state is $\Phi_{\text{target}}=|0\rangle_1  \frac{1}{\sqrt{2}} (|0\rangle_2 + |1\rangle_2)| \{0_k\}\rangle$. Equation \eqref{c2} indicates that the steady-state behavior of non-Markovian dynamics is critically determined by the number of bound states present in the energy spectrum of the total system. Specifically, the presence of one or two bound states causes $c_j(\infty)$ to stabilize at a constant value or exhibit sustained oscillations. These dynamical behaviors provide a robust feature for generating persistent entanglement and facilitating quantum state transfer between the SiV centers, thereby overcoming two principal limitations in quantum routing. This key phenomenon, which cannot be captured by the Markov approximation, crucially confirms its unique value for noise-resilient quantum routers.

\begin{figure}
\includegraphics[width=1\columnwidth]{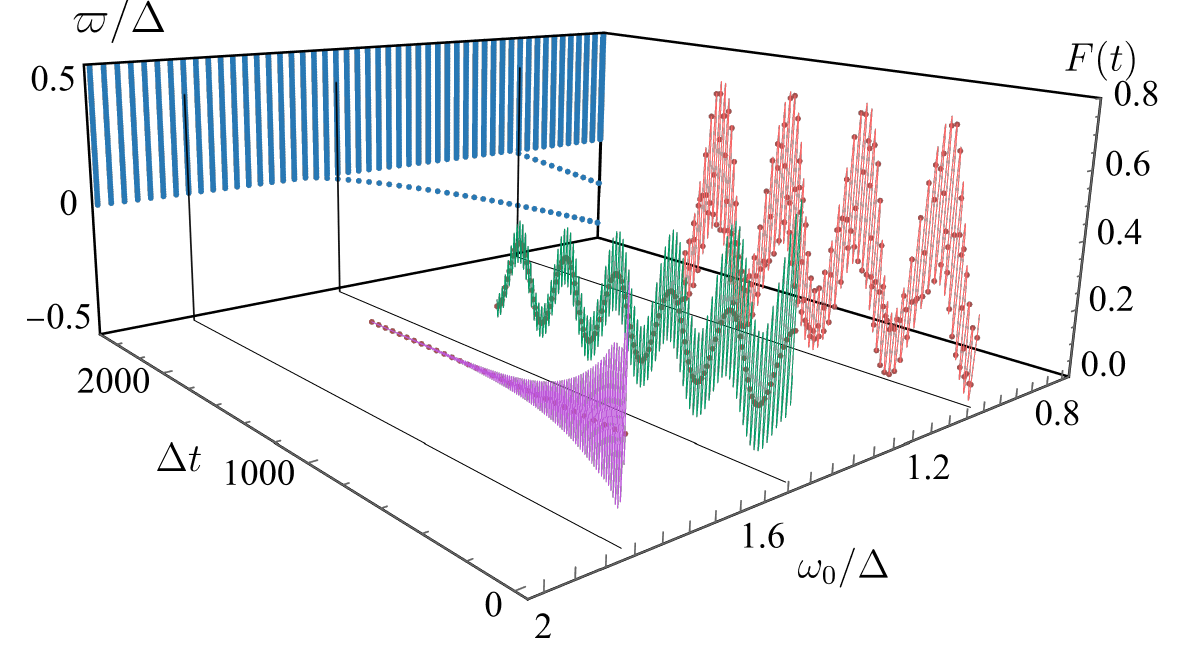}
\caption{Evolution of the state-transfer fidelity obtained from numerical solutions of Eq. \eqref{c} for different values of \(\omega_0\). The purple, green, and red curves denote the results with zero, one, and two bound states, respectively. The gray dots denote analytical results of Eq. \eqref{c2}, which show a good agreement with the numerical results. The parameters are the same as those in Fig. \ref{F2}.} \label{F3}
\end{figure}

We begin by examining the entanglement characteristics of the quantum router under the non-Markovian dynamics. As depicted in Fig. \ref{F2}, we show both the energy spectrum of the SiV-phonon system and the corresponding dynamics of concurrence \(C(t)\) at different SiV transition frequencies $\omega_0$. We find that in contrast to the complete decay to zero when $\omega_0=1.8\Delta$, \(C(t)\) asymptotically approaches a finite value when $\omega_0=1.525\Delta$. It is surprising to see that $C(t)$ even tends to a lossless Rabi-like oscillation when $\omega_0=1.05\Delta$. Being completely absent in the conventional Markov approximate dynamics, these diverse behaviors are well explained by the features of the energy spectrum. We observe that the two branches of bound states in the band gap divide the spectrum into three distinct regions. When $\omega_0> 1.58\Delta$, no bound state is formed and \(C(t)\) decays entirely to zero. When \(1.07\Delta <\omega_0 \leq1.58\Delta\), one bound state emerges, which leads to a finite steady-state entanglement. When \(\omega_0  \leq1.07\Delta\), the formation of two bound states results in persistent Rabi-like oscillations of \(C(t)\) in a frequency proportional to $|\varpi_1^b-\varpi_2^b|$. These diverse dynamical features excellently match the analytical predictions of Eq. \eqref{c2}. It demonstrates that the bound states and non-Markovian effects are crucial in sustaining entanglement for the quantum router. The result also reveals that by appropriately selecting the operating frequency $\omega_0$ of SiV centers, persistent quantum correlation is achievable for the distant SiV centers even though there is no direct interaction between them.

Our analysis has assumed identical SiV centers for the sake of clarity. However, practical devices inevitably exhibit a random distribution in their quantum properties. To assess the impact of the random distribution of the frequency of the $N$ SiV centers on our quantum-router scheme, we have performed a numerical calculation by adding a random frequency shift $\chi$ on the frequency of the SiV centers. As shown in Fig. \eqref{F2}(c), when the random frequency fluctuation is confined to a narrow range $[-0.5\Delta, 0.5\Delta]$, the averaged concurrence $C(t)$ remains close to the ideal case, indicating that the router can still function well. However, when the range of the frequency fluctuation increases to $[-\Delta, 5\Delta]$, the averaged $C(t)$ drops significantly and the fluctuations become large, implying a severe degradation in both entanglement generation and routing fidelity. Furthermore, when the $N$ SiV centers possess different frequencies, the quantitative condition for forming the bound states might be changed. However, as long as the bound states are present, the persistent entanglement between the SiV centers over long distances is achievable. These indicate that there is a clear tolerance to the parameter randomness for the reliable operation of our quantum-router scheme. The extremely large frequency expansion and mismatch of SiV centers would definitely lead to a decline in entanglement-generation efficiency, a reduction in quantum-state fidelity, and an increase in the unpredictability of system behavior. From an experimental perspective, it may be feasible to suppress these unwanted imperfections by exploiting static strain control to tune the frequencies of SiV centers \cite{sohn_controlling_2018}.

To demonstrate the state-transfer performance of the quantum router, we plot in Fig. \ref{F3} the time-dependent fidelity $F(t)$ between the two SiV centers. When $\omega_0> 1.58\Delta$, no bound state is present and \(F(t)\) decays to the theoretical minimum of $0.25$. In the regime \(1.07\Delta <\omega_0 \leq1.58\Delta\), a single bound state emerges, leading to Rabi-like oscillations in \(F(t)\) with a frequency proportional to $\varpi^b$. When \(\omega_0  \leq1.07\Delta\), two bound states are present, resulting in lossless oscillations with three distinct frequencies $\varpi_1^b$, $\varpi_2^b$, and $|\varpi_1^b-\varpi_2^b|$. We also observe a substantial improvement in \(F(t)\) compared to the Markov approximation, particularly in the presence of two bound states. All the results are in a good agreement with our analytical results in Eq. \eqref{fdtl}. These lossless oscillations demonstrate a robust transfer of quantum states between SiV centers mediated by the phonons. It establishes a stable platform for realizing the quantum router.

\begin{figure}
\includegraphics[width=1.0\columnwidth]{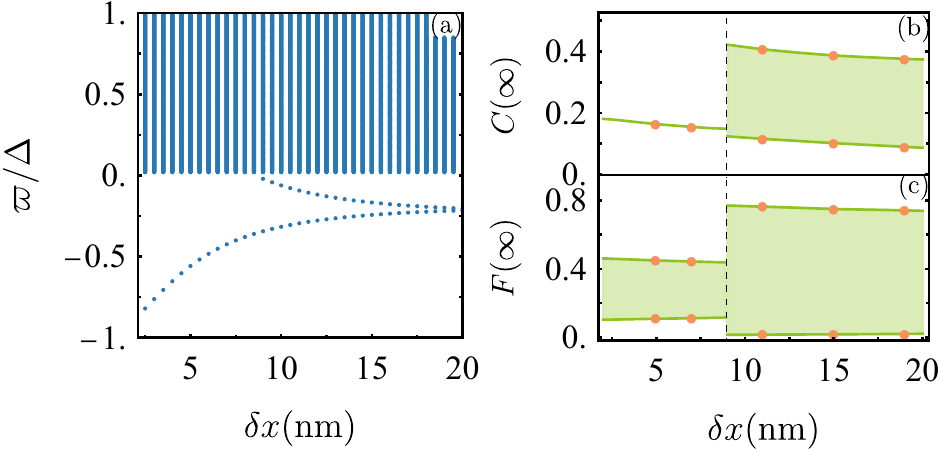}
\caption{(a) Energy spectrum, (b) long-time concurrence \(C(\infty)\), and (c) long-time state-transfer fidelity \(F(\infty)\) as a function of the inter-SiV distance $\delta x$ when \(\omega_0=\Delta\). In (b) and (c), the green curves indicate the maximum and minimum values derived from Eq. \eqref{c2} and the red dots denote the numerical results obtained from Eq. \eqref{c}. The green regions cover the values of \(C(\infty)\) and \(F(\infty)\) during their persistent oscillation. Other parameters are the same as Fig. \ref{F2}.}\label{F4}
\end{figure}

To further test the robustness of entanglement and state transfer in the quantum router against increasing inter-SiV distance, Fig. \ref{F4} shows the energy spectrum, the steady-state entanglement \(C(\infty)\), and the fidelity \(F(\infty)\) as a function of the distance \(\delta x\). When $\delta x< 9\mathrm{nm}$, only one bound state exists and thus \(C(\infty)\) stabilizes at a finite value and \(F(\infty)\) exhibits a periodic oscillation with its maxima and minima matching with the values evaluated from Eq. \eqref{c2}. When $9\mathrm{nm}\leq\delta x<20\mathrm{nm} $, two bound states appear. It causes both \(C(\infty)\) and \(F(\infty)\) to exhibit persistent oscillations. The one-to-one correspondence between the energy-spectrum features in Fig. \ref{F4}(a) and the dynamical behaviors in Figs. \ref{F4}(b) and \ref{F4}(c) demonstrates that the bound states play a key role in establishing a persistent interconnect between the SiV centers even when their distance is large. With further increasing the distance between the SiV centers, the eigenenergies $\hbar\varpi_i^b$ of the two bound states increasingly nearer. It causes the decrease of the oscillation frequency of the concurrence, which is proportional to $|\varpi_1^b-\varpi_2^b|$. However, our result that a persistent entanglement and suppressed decoherence of the SiV centers are achievable when two bound states are present in the energy spectrum of the total system formed by the SiV centers and the phonon waveguide does not change. This demonstrates the robustness of our quantum-router scheme to the separation of the SiV centers.

The scheme can be scaled up to multiple SiV centers. In the case of $N=3$, the spectral density matrix becomes $\mathbf{J}(\omega)=\begin{pmatrix} J_{0}(\omega) & J_{1}(\omega) & J_{2}(\omega) \\ J_{1}(\omega) & J_{0}(\omega) & J_{1}(\omega) \\ J_{2}(\omega) & J_{1}(\omega) & J_{0}(\omega) \end{pmatrix}$. We have $\mathbf{K}(s)=\operatorname{diag}\big[\tilde{h}_{0}-\tilde{h}_{2},\tilde{h}_{0}+\frac{1}{2}(\tilde{h}_{2}-\tilde{f}),\tilde{f}_{0}+\frac{1}{2}(\tilde{h}_{2}+\tilde{f})\big]$ and $\mathbf{A}_s = \begin{pmatrix} -1 & 1 & 1 \\ 0 & \dfrac{\tilde{h}_1(\tilde{f}-3\tilde{h}_2)}{\tilde{h}_2\tilde{f}-2\tilde{h}_1^2-\tilde{h}_2^2} & \dfrac{\tilde{h}_1(\tilde{f}+3\tilde{h}_2)}{\tilde{h}_2\tilde{f}+2\tilde{h}_1^2+\tilde{h}_2^2} \\ 1 & 1 & 1 \end{pmatrix} $, where the argument $ s $ in $ \tilde{h}_j(s) $ is omitted for brevity and $ \tilde{f} = \sqrt{8\tilde{h}_1^2 + \tilde{h}_2^2} $. We thus have $ \tilde{\mathbf{c}}(s) = \mathbf{Q}[s + i\omega_0 + \mathbf{K}(s)]^{-1} \begin{pmatrix} 1 & 1 & 1 \end{pmatrix}^T $, where $ \mathbf{Q} = \begin{pmatrix} 1/2 & \dfrac{1-\tilde{h}_2/\tilde{f}}{4} & \dfrac{1+\tilde{h}_2/\tilde{f}}{4} \\ 0 & -\tilde{h}_1/\tilde{f} & \tilde{h}_1/\tilde{f} \\ -1/2 & \dfrac{1-\tilde{h}_2/\tilde{f}}{4} & \dfrac{1+\tilde{h}_2/\tilde{f}}{4} \end{pmatrix} $. By applying the residue theorem, we obtain
\begin{equation}
c_j(\infty) = \sum_{l=1}^{3} Q_{jl} Z_l e^{-i\varpi_l^bt}.\label{c3}
\end{equation}
 \begin{figure}
    \includegraphics[width=1\columnwidth]{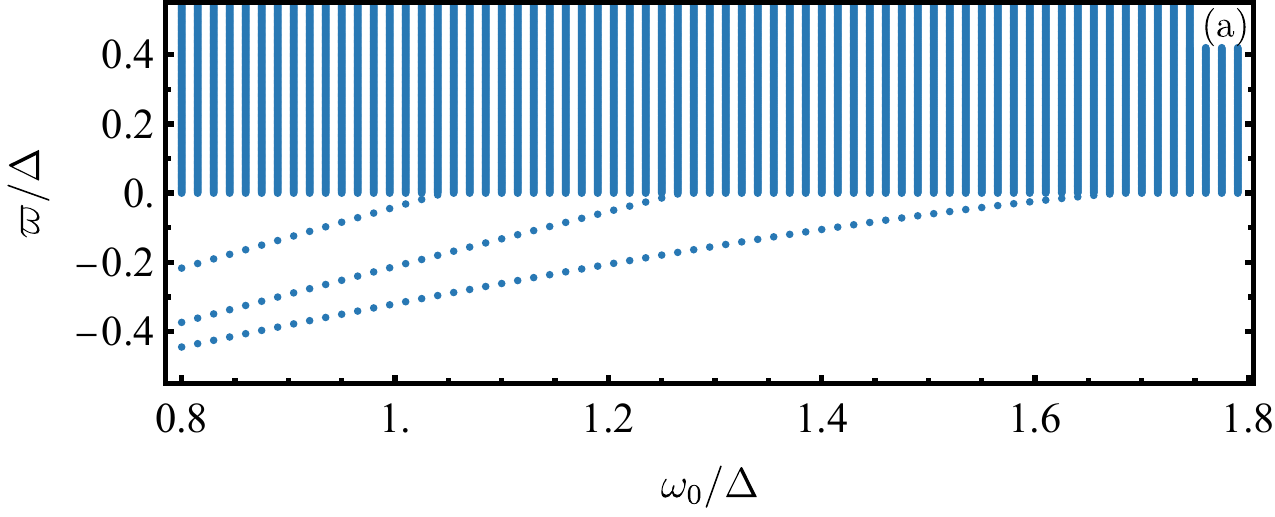}
        \includegraphics[width=1\columnwidth]{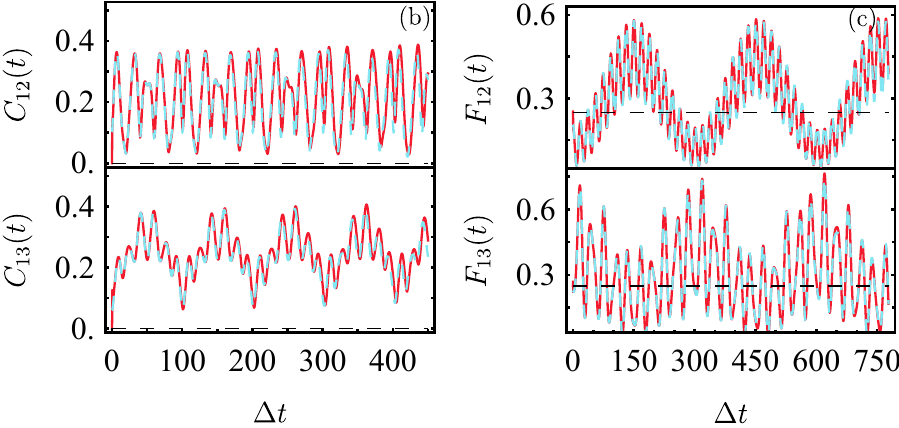}
	\caption{(a) Energy spectrum of total system. (b) Time evolution of bipartite entanglement \(C_{12}(t)\) and \(C_{13}(t)\). (c) Time evolution of the fidelity \(F_{12}(t)\) and \(F_{13}(t)\). In (b) and (c), the cyan dashed curves correspond to results from Eq. \eqref{c}, while the red solid curves are derived from Eq. \eqref{c3}, both for the case with three bound states when \(\omega_0=\Delta\). For comparison, the black dashed lines from Eq. \eqref{c3} show the case without bound states when \(\omega_0=2\Delta\). The parameters are the same as in Fig. \ref{F2} except for $\delta x = 10.5\mathrm{nm}$.}\label{F5}
\end{figure}
\begin{figure}
	\includegraphics[width=1.0\columnwidth]{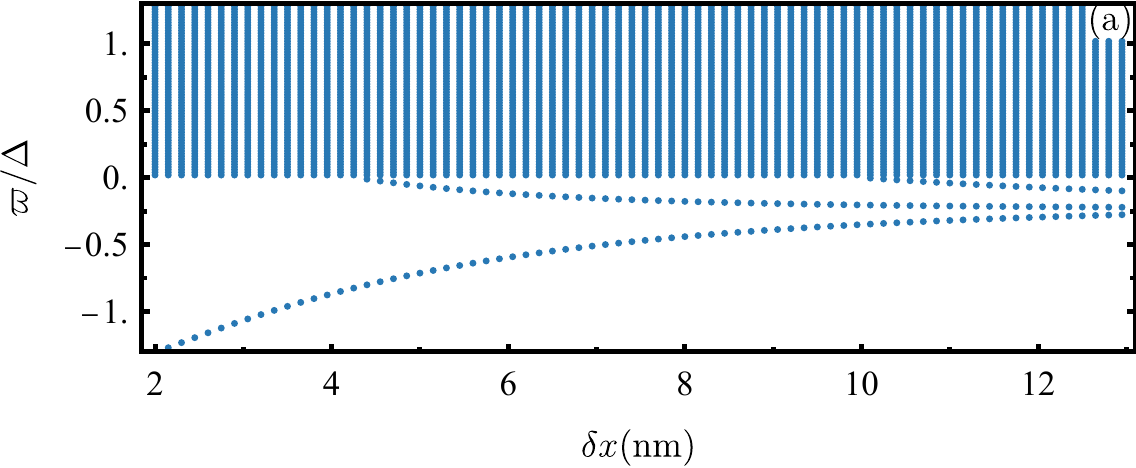}
    	\includegraphics[width=1.0\columnwidth]{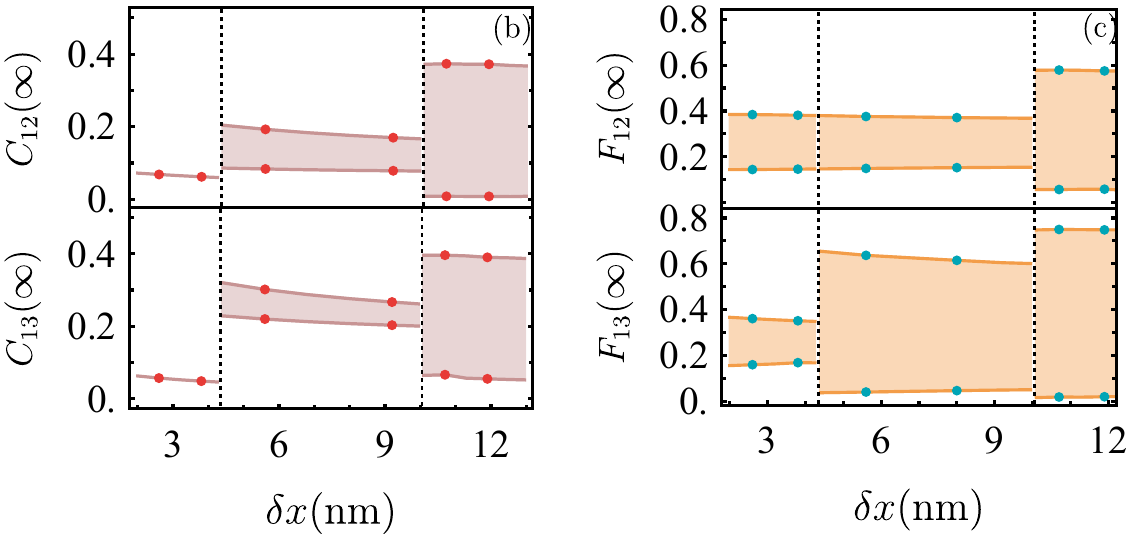}
 \caption{(a) Energy spectrum versus distance $\delta x$ at \(\omega_0=\Delta\). (b) Steady-state entanglement \(C_{12}(\infty)\) and \(C_{13}(\infty)\) exhibits oscillations within the purple-shaded region. The purple lines represent the maximum and minimum values of steady-state entanglement obtained from Eq. \eqref{c3}, while the red dots represent numerical results from Eq. \eqref{c} at selected distances. (c) Corresponding state transfer fidelity \(F_{12}(\infty)\) and \(F_{13}(\infty)\) oscillate within the orange-shaded region, with their extrema determined by Eq. \eqref{c3} (orange lines). Green dots indicate numerical solutions from Eq. \eqref{c} at specific $\delta x$ values. Other parameters are the same as Fig. \ref{F2}.}\label{F6}
\end{figure} 

Figure \ref{F5}(a) shows three bound-state branches emerging in the energy spectrum of the three-node quantum router. In the absence of bound states, decoherence causes the entanglement between any bipartite partition to decay to zero and the fidelity between SiV$_1$ and SiV$_2$/SiV$_3$ nodes to decay to its lower limit of $0.25$; see Figs. \ref{F5}(b) and \ref{F5}(c). When three bound states are present, SiV$_1$ preserves  simultaneous entanglement with SiV$_2$ and SiV$_3$ in the quantum router. The entanglement in this case exhibits trifrequency oscillations, a behavior clearly demonstrated by the red solid and green dashed curves in Figs. \ref{F5}(b) and \ref{F5}(c). This configuration also significantly improves the state-transfer fidelity from SiV$_1$ to SiV$_2$/SiV$_3$, with its dynamics displaying complex multifrequency oscillations. The exact match between the analytical (red solid) and numerical (blue dashed) results in Figs. \ref{F5}(b) and \ref{F5}(c) conclusively validates the essential role of bound states in the quantum router. Similarly to the $N=2$ case, the three-SiV quantum router maintains quantum coherence over long distances and exhibits the spatial robustness. As shown in Fig. \ref{F6}, the energy spectrum is presented along with the steady-state values of \(C_{12}(\infty)/C_{13}(\infty)\) and \(F_{12}(\infty)/F_{13}(\infty)\) as a function of the distance \(\delta x\). The results demonstrate that under the protection of bound states, both the entanglement and the fidelity in the quantum router can maintain either stable finite values or persistent multifrequency oscillations even at long-distance node separations, revealing remarkable decoherence-resistant properties. These results demonstrate that the formation of bound states not only can sustain persistent entanglement among multiple nodes but also realize parallel one-to-many quantum-state transfer. Thus, it establishes a robust noise-resistant quantum router for multinode networks.

\section{Discussion and conclusion}\label{3}

Our solid-state quantum router is realizable under state-of-the-art nanofabrication techniques, including advanced strain control and precision ion implantation. It maintains complete compatibility with current experimental platforms \cite{Toyli2010,PhysRevApplied.7.064021}. As demonstrated in Ref. \cite{sohn_controlling_2018}, three arrays of SiV centers are successfully integrated into a diamond cantilever at precisely designed positions. In general, the proposed setup can be experimentally realized by embedding an array of SiV centers into a one-dimensional diamond waveguide. The spacing between SiV centers embedded in diamond waveguides in this paper can reach 20 $\mathrm{nm}$, which is significantly smaller than the phonon wavelength. Nevertheless, recent studies have demonstrated that strong phonon correlations can indeed be established and harnessed for efficient heat transport even in nanostructures with periodic features smaller than the phonon wavelength \cite{Xie01012018,https://doi.org/10.1002/qute.202100074}. Moreover, the spatial precision in fabricating SiV centers has been reported at the sub-30-$\mathrm{nm}$ level \cite{bradac_quantum_2019}, further supporting the feasibility of achieving such small intercenter distances in diamond waveguides. Although achieving precise positional control of SiV centers and fabricating waveguides with cross-sectional areas as small as 100 $\mathrm{nm}^2$ remains challenging with current technologies, continued progress in diamond nanostructure fabrication is expected to address these limitations. The diamond waveguide, with material parameters \( \varrho = 3500 \, \text{kg/m}^3 \), \( E = 1050 \, \text{GPa} \), and \( \nu = 0.2 \), supports phonon propagation at a velocity \( v \approx 1 \times 10^4 \, \text{m/s} \). These satisfy the requirement of our scheme. As a final remark, our scheme also works if the initial quantum information is encoded in another SiV center instead of the first one. This inherent flexibility is another advantage of our router architecture: any node in the network can act as the input port, enabling dynamic and reconfigurable quantum information routing based on need, rather than being limited to a fixed and pre-determined path.

In summary, we have proposed a quantum router based on SiV centers coupled to phonon modes in a one-dimensional diamond waveguide. We have discovered that the emergence of bound states in the energy spectrum of the total system enables persistent quantum entanglement and facilitates noise-resilient state transfer among distant SiV centers. This mechanism allows the realization of a multinode quantum router without the need for complex external controls. By harnessing the bound-state mechanism, we have designed a scalable non-Markovian quantum router that simultaneously solves the challenges of distance limitations, fidelity degradation, and parallel interconnections faced by conventional quantum routers. Our finding establishes an engineered solid-state system as a viable platform for quantum routers and opens an avenue for developing quantum networks.

\begin{acknowledgements}
The work is supported by the National Natural Science Foundation of China (Grants No. 12275109, No. 92576202, and No. 12247101), the Quantum Science and Technology-National Science and Technology Major Project (Grant No. 2023ZD0300904), the Fundamental Research Funds for the Central Universities (Grant No. lzujbky-2025-jdzx07), and the Natural Science Foundation of Gansu Province (No. 22JR5RA389 and No. 25JRRA799).
\end{acknowledgements}

\bibliography{ref}  

\end{document}